%% file: main.tex
\documentclass{llncs}
\usepackage[utf8]{inputenc}

\title{Everything You Always Wanted to Know About TREC RTS* \\{\small(*But Were Afraid to Ask)}}
\author{Gilles Hubert, Jose G. Moreno, Karen Pinel-Sauvagnat and Yoann Pitarch}
\institute{IRIT, University of Toulouse, Toulouse, France\\
\{gilles.hubert, jose.moreno, karen.sauvagnat, yoann.pitarch\}@irit.fr}
\date{}

\usepackage{graphicx}
\usepackage{booktabs}
\usepackage{comment}
\usepackage{amssymb}
\usepackage{pifont}
\usepackage{array}
\usepackage{multirow}
\usepackage{comment}
\usepackage{soul, color}
\usepackage{todonotes}
\usepackage{url}
\usepackage{float}
\usepackage{caption}
\usepackage{mathtools}

\newcommand{\overbar}[1]{\mkern 1.5mu\overline{\mkern-1.5mu#1\mkern-1.5mu}\mkern 1.5mu}

\begin{document}

\maketitle

\begin{abstract}
The TREC Real-Time Summarization (RTS) track provides a framework for evaluating systems monitoring the Twitter stream and pushing tweets to users according to given profiles. It includes metrics, files, settings and hypothesis provided by the organizers. In this work, we perform a thorough analysis of each component of the framework used in 2016 and 2017 and found some limitations for the Scenario A of this track. Our main findings point out the weakness of the metrics and give clear recommendations to fairly reuse the collection.

\end{abstract}

\input{sections/intro}

\input{sections/overview}

\input{sections/metrics}

\input{sections/discussionProtocole}

\input{sections/hypotheses}

\input{sections/discussion}

\input{sections/rerun}

\input{sections/reco}

\bibliographystyle{splncs03}
\bibliography{references}
\end{document}

%% file: sections/intro.tex
\section{Introduction}

A common usage of Twitter is to watch other users' tweets and never post anything. This usage scenario considers Twitter as a real time information source by scanning incessantly the tweet stream. 
The users adopting this usage aim to catch new (information they did not hear about before), fresh (information that appeared very recently) and precise (information that concerns them) information. There is growing interest in systems that could address these issues by providing information that satisfy this type of users with respect to their information needs.

The TREC campaign took an interest in the evaluation of such systems through various tracks and notably the ongoing Real-Time Summarization (RTS) track.
As usual in information retrieval and evaluation campaigns, the researchers who tackle the issues on which focuses a track test their approaches using the framework provided for the track during the campaign period. Many researchers also test their solutions using the framework after the campaign period.

Our participations to these successive tracks have motivated a thorough analysis of the provided evaluation frameworks. 
This paper presents our main findings about the scenario A of the 2016 and 2017 benchmarks. On the one hand, it highlights some limitations of the provided evaluation framework with respect to the organizers' settings. On the other hand, it identifies precautions to take when reusing the evaluation framework after the campaign period. Leaving aside these precautions would lead to erroneous evaluation results and invalidate conclusions on system performance.  

The remainder of this paper is organized as follows. Section~\ref{sec:overview} presents an overview of the TREC RTS track while Section~\ref{sec:metrics} describes the metrics defined for the evaluations of systems corresponding to the track scenario A. The limitations highlighted on the evaluation framework are discussed in Sections~\ref{sec:hypotheses} and \ref{sec:hyperparameters}. Section~\ref{sec:rerun} introduces the precautions to take to obtain valid results when reusing the evaluation framework after the TREC campaign. Finally, Section~\ref{sec:reco} concludes the paper.

%% file: sections/overview.tex
\section{Overview of the TREC RTS Scenario A}\label{sec:overview}

Introduced in 2016 and continued in 2017, the RTS track merges some previous TREC tracks: the Microblog (MB) track ran from 2010 to 2015 and the Temporal Summarization (TS) track run from 2013 to 2015. It intends to promote the development of systems that automatically monitor a document stream to keep the user up-to-date on topics of interest, by proposing a framework to evaluate such systems. 
The track considers two scenarios: scenario A -- Push notifications -- and scenario B -- Email digest.
The scenario A corresponds to the systems intended to send immediately the posts identified as relevant. The scenario B corresponds to systems intended to send once a day a summarization of the relevant posts of the day.
This paper sheds some light on the evaluation framework defined for the scenario A.

\begin{table}[H]
\centering

{\scriptsize
\setlength{\tabcolsep}{0.16cm}
\begin{tabular}{llcc}\\\toprule
Year&Evaluation period& \# judged topics&\# competitors \\ \midrule
2016&From 02/08/2016 00:00:00 to 11/08/2016 23:59:59 &56&41 \\
2017&From 29/07/2017 00:00:00 to 05/08/2017 23:59:59&97 &41 \\\bottomrule
\end{tabular}}
\caption{Statistics of scenario A in 2016 and 2017. Times are provided in UTC in order to have fixed time intervals regardless the participant location.}
\label{tab:statTask}
\end{table}

 Each participant to the task must process a publicly accessible sample provided by Twitter which corresponds to the 1\,\% of the total available tweets.
The evaluation period is partitioned in days, making 8 or 10 days long the evaluation window. 
To identify relevant tweets, a set of profiles is provided. Each profile (called \textit{topic} in the TREC jargon) is composed of a title, a description and a narrative of the interest profiles. Table~\ref{tab:statTask} provides some statistics about the 2016 and 2017 tasks.
Each system must push at most 10 tweets per profile per day to a central system called the broker. Note that silence of a system is a desired effect when there are no relevant tweets during a day.

Two ways of evaluation were performed: online judgments and batch judgments. The earlier was performed during the evaluation period and the latter was performed once the challenge was over. Some works studied these two ways of evaluation and showed they are correlated \cite{Roegiest2017,Tan2016}. This work is interested only in the latter due to the reusability problems already identified in the earlier one \cite{TanSigir2017noreusability}.
In order to perform the batch judgments, a pool of tweets was built using all the pushed tweets in both scenarios A and B. The combined set of tweets was annotated following a two-step methodology. Given a tweet, assessors first assigned a relevance score. In all editions of this task, three levels of relevance (not relevant, relevant and very relevant) were considered. However without loss of generality, we consider only two levels of relevance to simplify our study, i.e., relevant tweets are considered as very relevant.
Then, a unique cluster\footnote{A cluster can be considered as a group of tweets sharing the same semantic information.} was assigned to each relevant tweet. A tweet is considered relevant if its content is related to one profile.  The clusters were found following the Tweet Timeline Generation (TTG) approach~\cite{WangSigir2015TTG} which takes into account the creation timestamp to sort relevant tweets. Tweets are examined one per one traveling from past to future. A new cluster is created if the current tweet content is substantially dissimilar to all the previous tweets seen. All clusters are then considered equally important in the evaluation metrics.

%% file: sections/metrics.tex
\section{Metrics}\label{sec:metrics}


The RTS track in its guidelines asks for effectiveness (tweet quality) and efficiency (no latency).
As participant systems might favour effectiveness or efficiency depending on their approaches, the organizers decided from 2016 to compute metrics for quality and latency separately \cite{lin2016}.

\input{sections/notations}

\subsection{Gain-Oriented Metrics}

\subsubsection{Gain.}
The three metrics proposed to evaluate quality are based on the concept of gain, i.e., the  usefulness of a tweet in the list of the tweets pushed by the system. The way the gain is evaluated is thus decisive. 
Given a time window $w_j$ and $T_i(w_j)$, i.e., the tweets returned by the system $S_i$ published during $w_j$, the gain $G(w_j, S_i)$ is evaluated as follows:
\begin{equation}
G(S_i, w_j)= \sum_{t \in T_i(w_j)} g(t) 
\label{eq:gain}
\end{equation}
where $g(t)$ is the gain of the tweet $t$: $g(t)=1$ if the tweet is relevant, $g(t)=0$ otherwise, i.e., $t$ is non relevant or redundant. It should be noted that this definition has been clarified from \cite{trec2017,lin2016} by specifying that the tweets considered during $w_j$ are picked using $\Theta(t)$ rather than $\Pi_i(t)$. We now detail the official metrics that rely on the gain.

\subsubsection{Expected gain.}
The expected gain metric, denoted by \textit{EG}, is adapted from \cite{aslam2015}. Given a time window $w_j$, it is evaluated as:
\begin{equation}
\mathit{EG} (w_j,S_i) = \frac{1}{|T_i(w_j)|} \cdot G(S_i, w_j)
\end{equation}
where $|T_i(w_j)|$ is the number of tweets  returned by $S_i$ and published during $w_j$.

An important question about this metric is how to score systems during the so-called silent days, i.e., the days where no relevant tweets are published.
Some variants of the \textit{EG} metric have been introduced differing on how the silent days are considered:
\begin{itemize}
\item \textit{EG}-0 in which systems receive a gain of 0 during the silent days no matter the tweets they returned.
\item \textit{EG}-1 in which systems receive a gain of 1 during the silent days when they do not return any tweet published during the day, 0 otherwise. It should be noted that this definition has been slightly extended from \cite{lin2016} to perfectly fit with the evaluation tool. This will be further discussed in Section~\ref{sec:hypotheses}.
\item \textit{EG}-p in which the proportion of tweets returned during a silent day is considered: a system receives a score of $\frac{N- |\overbar{t}|}{N}$, where  $|\overbar{t}|$ is the number of non-relevant tweets published during the day and returned by the system. For instance, if a system pushes one tweet published during the day but not relevant (instead of 0), it gets a score of 0.9; two non-relevant tweets imply a score of 0.8, etc. Similarly to \textit{EG}-1, this definition has been slightly extended.
\end{itemize}

The way the silent days are considered is crucial, since a huge impact of silent vs. eventful days is observed in the evaluation
\cite{Tan2016}.

\subsubsection{Normalized Cumulative Gain.}

Given a time window $w_j$, the \textit{nCG} metric is evaluated as follows:
\begin{equation}
\mathit{nCG} (w_j,S_i) = \frac{1}{\cal Z} \cdot G(S_i, w_j)
\end{equation}

$\cal Z$ is the maximum possible gain (given the $N$ tweets per day limit).
As for $EG$, three variants are considered regarding how the silent days are taken into account: \textit{nCG}-1, \textit{nCG}-0, and \textit{nCG}-p.

\subsubsection{Gain Minus Pain.}
The \textit{GMP} metric evaluates the utility of the run:
\begin{equation}
\mathit{GMP}(S_i, w_j)  = \alpha \sum G(S_i, w_j)  - (1 - \alpha) \cdot P(S_i, w_j)
\end{equation}

The gain Gain $G(S_i, w_j)$ is computed in the same manner as above,
the pain $P(S_i, w_j)$ is the number of non-relevant tweets published during $w_j$ and returned by the system $S_i$, and $\alpha$ controls the balance between the two. Three $\alpha$ settings were considered: 0.33, 0.50, and 0.66.

\subsection{Latency-Oriented Metric}
The latency metric is defined as:

\begin{equation}
\mathit{Latency}(S_k)=\sum_{t_i^{(\cdot)} \in R_k} \Pi_k(t_i^{(\cdot)}) - \Theta(t_i^{1})
\end{equation}
 where $t_i^{(\cdot)}$ is the oldest tweet pushed by the system $S_k$ for the cluster $C_i$.
 
In other terms, latency is evaluated only for tweets contributing to the gain as the difference between the time a tweet was pushed and the first tweet in the semantic cluster that the tweet belongs to.
 
\subsection{Metrics Exemplifications}

\begin{figure}[t]
  \centering
  \includegraphics[width=0.7\textwidth]{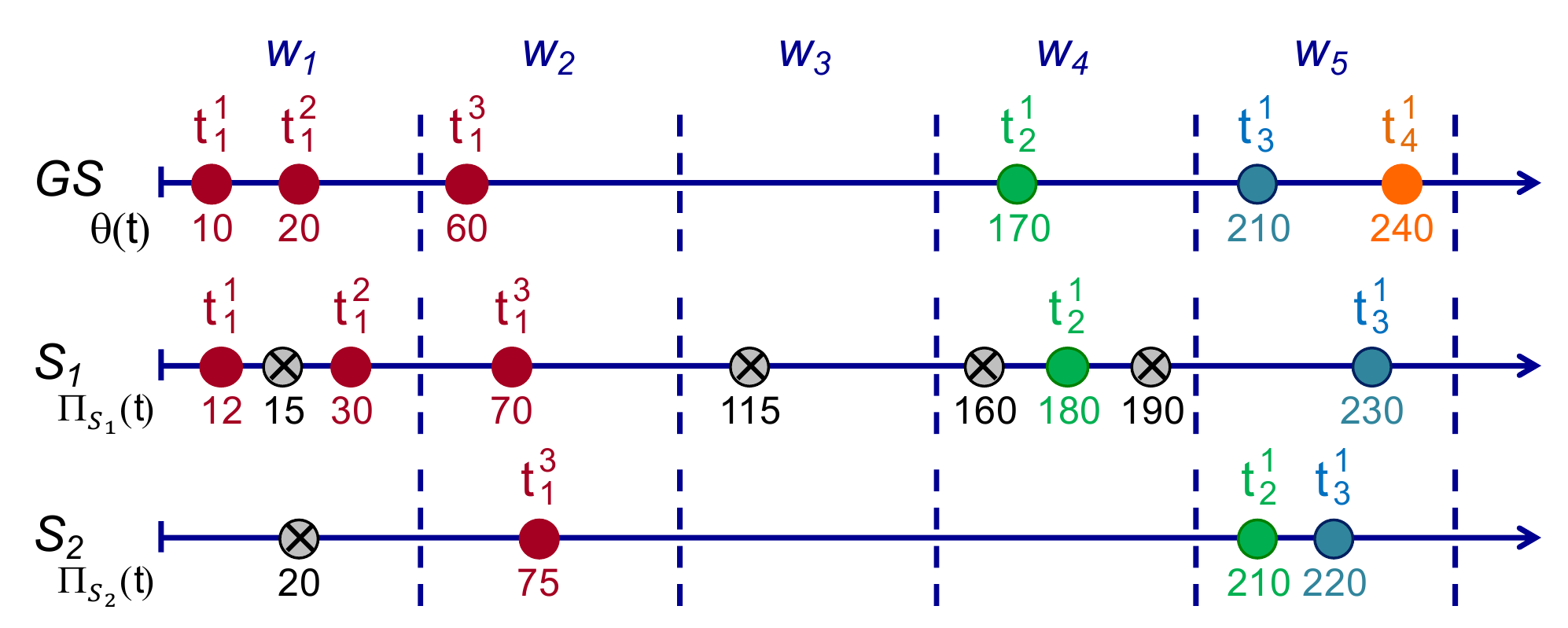}
  \caption{Examples of runs retrieved by $S_1$ and $S_2$ as well as the associated ground truth ($GS$). In this example, a time window $w_i$ lasts 50 seconds. }
 \label{fig:examples} 
\end{figure}

Fig. \ref{fig:examples} and Table \ref{tab:examples} run through some examples of systems and the way the official metrics are calculated. 
The results presented in Table \ref{tab:examples} are decomposed and were checked using the 2016 and 2017 official evaluation tools of the track\footnote{Official evaluation tools are available at \url{http://trec.nist.gov/data/rts2016.html} (2016) and \url{http://trec.nist.gov/act_part/act_part.html} (2017), last checked: October 6, 2017.}. 

In the \textit{EG} family of metrics, the gain in a time window is divided by the number of tweets returned by the system and published during the time window. For instance and considering the system $S_1$, the gain in the time window $w_1$ is divided by 3 while it is divided by 1 in $w_2$.
In the \textit{nCG} family of metrics, the gain in a time window is divided by the optimal gain. For instance, in the time window $w_5$ the optimal gain is 2 (2 new clusters $C_3$ and $C_4$), but neither $S_1$ nor $S_2$ reach this optimal gain.  

If we now consider the silent days, \textit{EG}-1 and \textit{nCG}-1 reward the systems for returning no tweets, and strongly penalize them otherwise. For instance, $S_1$ breaks the silence during $w_3$ and thus obtains a score of $0$ for this window.
The silent days can be different from one system to another: 
$S_1$ breaks the silence during $w_2$ since $t_1^3$ is not relevant in this case (the $C_1$ cluster has already been retrieved), whereas   this is not the case for $S_2$ for which $C_1$ was not retrieved at this time.
Conversely, $S_1$ and $S_2$ receive a score of $0$ for $w_3$ considering the \textit{EG}-0 and \textit{nCG}-0 metrics, whereas $S_2$ has a perfect behavior during this window. Whatever the systems return, the silent days are associated with a score of $0$, and it never hurts to push tweets. 
The ``silent days effect" is lowered for the evaluation of \textit{EG}-p et \textit{nCG}-p: $S_1$ receives a score of $9/10$ on $w_2$ and $w_3$ (whereas it receives a score of 0 for \textit{EG}-0, \textit{nCG}-0, \textit{EG}-1, and \textit{nCG}-1 metrics).

\begin{table}[t]
\centering

\tiny
\begin{tabular}{lp{0.45\textwidth}p{0.45\textwidth}}\\\toprule
Metrics         & \multicolumn{2}{c}{Systems}\\\cmidrule{2-3}
         & \multicolumn{1}{c}{$S_1$} & \multicolumn{1}{c}{$S_2$}\\\midrule
\textit{EG}-0         &  $(\frac{1}{3}*1+0+0+\frac{1}{3}*1+\frac{1}{1}*1)/5=0.33$ & $(0+\frac{1}{1}*1+0+\frac{1}{1}*1+\frac{1}{1}*1)/5=0.6$ \\[0.10cm]
\textit{EG}-1         & $(\frac{1}{3}*1+0+0+\frac{1}{3}*1+\frac{1}{1}*1)/5=0.33$ & $(0+\frac{1}{1}*1+1+\frac{1}{1}*1+\frac{1}{1}*1)/5=0.8$ \\[0.10cm]
\textit{EG}-p        & $(\frac{1}{3}*1+\frac{9}{10}+\frac{9}{10}+\frac{1}{3}*1+\frac{1}{1}*1)/5=0.69$ & $(0+\frac{1}{1}*1+1+\frac{1}{1}*1+\frac{1}{1}*1)/5=0.8$ \\[0.10cm]\midrule[0.03em]
\textit{nCG}-0        & $(\frac{1}{1}*1+0+0+\frac{1}{1}*1+\frac{1}{2}*1)/5=0.5$ & $(0+\frac{1}{1}*1+0+\frac{1}{1}*1+\frac{1}{2}*1)/5=0.5$ \\[0.10cm]
\textit{nCG}-1        & $(\frac{1}{1}*1+0+0+\frac{1}{1}*1+\frac{1}{2}*1)/5=0.5$ & $(0+\frac{1}{1}*1+1+\frac{1}{1}*1+\frac{1}{2}*1)/5=0.7$ \\[0.10cm]
\textit{nCG}-p       & $((\frac{1}{1}*1+\frac{9}{10}+\frac{9}{10}+\frac{1}{1}*1+\frac{1}{2}*1)/5=0.86$ & $(0+\frac{1}{1}*1+1+\frac{1}{1}*1+\frac{1}{2}*1)/5=0.7$ \\[0.10cm]
\midrule[0.03em]
\textit{GMP}.50       & $((0.5*1-0.5*2)+(0)+(-0.5*1)+(0.5*1-0.5*2)+(0.5*1))/5=-0.1$ &  $((-0.5*1)+(0.5*1)+(0)+(0.5*1)+(0.5*1))/5=0.2$ \\[0.10cm]
\midrule[0.03em]
\textit{Latency}     & $2+10+20=32$ & $65+40+10=115$\\\bottomrule
\end{tabular}
\small
\caption{Behaviours of the studied metrics with respect to the metrics.}
\label{tab:examples}
\end{table}

For all the gain-oriented metrics, a tweet participates to the gain of the time window on which it was published (and not on which it was pushed by the systems). As a consequence, if we consider the system $S_2$ and the window $w_5$, the $t^1_2$ tweet participates to the gain of $w_4$ (which is the window in which it was published). $S_2$ has thus a non-zero score for $w_4$ whereas it did not return any tweet.    
Another point to discuss relates to redundant tweets. As expected, $t_1^2$ returned by $S_1$ during $w_1$ is considered as not relevant, only $t_1^1$ participates to the gain.

At last, the \textit{Latency} metric is evaluated independently of the time windows as the difference between the first tweet found in the cluster by the system and the publication date of the first tweet in the cluster in the gold standard. A side effect of this metric is that a perfect latency can be obtained without returning any relevant tweets.

\subsection{Metric Integration in the Evaluation Framework}

\begin{table}
\centering

\scriptsize
\setlength{\tabcolsep}{0.16cm}
\begin{tabular}{lllccccc}\\[-0.5cm]\toprule
Metrics & Variants & Years & Recall & Precision & Utility & Latency & Averaged over \\ \midrule
\multirow{3}{0pt}{\textit{EG}} & \textit{EG}-1 & 2016*, 2017 &  & \ding{52}  &   &   & profiles  \\ 
        & \textit{EG}-0 & 2016 &  & \ding{52}  &   &  &   and  \\ 
        & \textit{EG}-p & 2017* &  & \ding{52}  &   &  &  days  \\ \midrule[0.03em]
\multirow{2}{0pt}{\textit{nCG}}      & \textit{nCG}-1 & 2016, 2017  &  \ding{52} &   &   &    & profiles \\ 
        & \textit{nCG}-0 & 2016  &  \ding{52} &   &    &  &  and  \\
        & \textit{nCG}-p & 2017 &  \ding{52} &   &    &   & days \\ \midrule[0.03em]
\multirow{3}{0pt}{\textit{GMP}} & \textit{GMP}.33 & 2016, 2017 &  &   & \ding{52} &  & profiles  \\
     & \textit{GMP}.50 & 2016, 2017 &  &   & \ding{52} &  &  and  \\ 
     & \textit{GMP}.66 & 2016, 2017 &  &   & \ding{52} &   & days \\ \midrule[0.03em]
Latency &    & 2016, 2017 &  &   &    & \ding{52} & profiles \\\bottomrule\\
\end{tabular}
\caption{Official metrics for the 2016 and 2017 tracks. The primary metric for each year is denoted with *.\label{tab:metrics}}
\end{table}

Table \ref{tab:metrics} provides additional information about the metrics. 
The \textit{EG} metrics attempt to capture precision while the \textit{nCG} ones are recall-oriented. The \textit{GMP} metrics aim to fill the gap between these two contradictory objectives and thus represent a trade-off between precision and recall. As stated in \cite{Roegiest2017}, \textit{EG}-0 and \textit{nCG}-0 metrics are poorly formulated metrics and were thus abandoned in 2017. 
The gain-oriented metrics are computed for each interest profile and each window $w_j$. The score for a competitor is the mean of the scores for each day over all the profiles. Since each profile contains the same number
of days, there is no distinction between micro- vs. macro-averages.
The \textit{EG}-1 and \textit{EG}-p metrics were respectively considered as the official metrics in 2016 and 2017.

%% file: sections/notations.tex
\subsection{Notations and Preliminary Definition}
The notations used in this paper are summarized in Table~\ref{tab:notations}.

\begin{table}[H]
\centering
\setlength{\tabcolsep}{0.16cm}

\begin{tabular}{lp{0.7\textwidth}}\\[-0.65cm]\toprule
  Notation 		& Definition\\\midrule
 $C=\{C_1, \ldots, C_k\}$ & The set of clusters \\
 $t_i^j$ & The $j^{th}$ tweet belonging to the cluster $C_i$ \\
 $\overbar{t}$ & A non relevant tweet \\
 $S_i$ & A system \\
 $\Theta(t)$ & The creation date of the tweet $t$ \\
 $\Pi_i(t)$ & The date at which the tweet $t$ has been pushed to the user by the system $S_i$\\
 $W=\{w_1, \ldots, w_T\}$ & The set of temporal windows, i.e., the set of days considered during the evaluation campaign\\
 $N$ & The maximum number of tweets to push per  window \\
 $T_i(w_j)$ & The list of tweets published during the window $w_j$ and pushed by the system $S_i$  ordered by their $\Theta(t)$ \\
 $R_i$ & The set of relevant tweets pushed by the system $S_i$\\

\bottomrule
\end{tabular}
\caption{Notations used throughout the paper}
\label{tab:notations}
\end{table}

A key concept for all the metrics is how relevance is defined. A tweet is considered as relevant for a system $S_i$ if it satisfies two criteria: it is contained in a relevant cluster and it is the first tweet returned by $S_i$ for this cluster. Once a tweet from a cluster has been retrieved, all the other tweets from the same cluster are redundant and automatically become not relevant \cite{lin2015}.
This implies that the relevance of a tweet is system-dependent.

%% file: sections/discussionProtocole.tex
\section{Hypotheses and Settings of the Evaluation Framework}
\label{sec:discussionProtocol}

The metric evaluation is based on two  hypotheses assumed by the organizers.\\

\noindent \textbf{H1 -- Redundant information is non relevant.} As mentioned in \cite{lin2016}: \begin{quote}\textit{
Once a tweet from a cluster is retrieved, all other tweets from the same cluster automatically become not relevant. This penalizes systems for returning redundant information.}\end{quote}

\noindent \textbf{H2 -- A perfect daily score is obtained when silence is respected}. As mentioned in \cite{lin2016}:  \begin{quote}
\textit{
In the EG-1 and nCG-1 variants of the metrics, on a ``silent day'', the system receives a score of one (i.e., perfect score) if it does not push any tweets, or zero otherwise.} \end{quote}
Note that since relevance is system-dependent, it implies that the silent days are system-dependent as well.
Moreover, we would like to shed some light on two settings of the framework.\\

\noindent{\bf S1 -- \mathversion{bold}{$N=10$}.} This consists in forcing the systems to  push a maximum of only 10 tweets per day and per profile. There is a twofold explanation for the value chosen for this parameter: first, to impose to the systems a realistic limit to the number of daily tweets that could be desired by a user and second, to impose a reasonable limit for the annotation phase. \\

\noindent{\bf S2 -- Evaluation window.}  For the gain-oriented metrics, whatever the $\Pi_i(t)$ value for a tweet, only $\Theta(t)$ is considered for the evaluation of $G(\cdot,\cdot)$ (see equation \ref{eq:gain}). In other terms, each returned tweet is sent back to its emission window. This can affect the systems that use buffering-based strategies as suggested by the guidelines. This setting is implicitly defined by the organizers since the latency metric is calculated separately from the main metrics.

\label{sub:defhypotheses}

%% file: sections/hypotheses.tex
\section{Metrics Adequacy under RTS Hypotheses}
\label{sec:hypotheses}
We now refute the aforementioned hypotheses through 2 counterexamples.

\begin{figure}
\begin{minipage}[b]{.35\textwidth}\centering
  \includegraphics[width=0.8\textwidth]{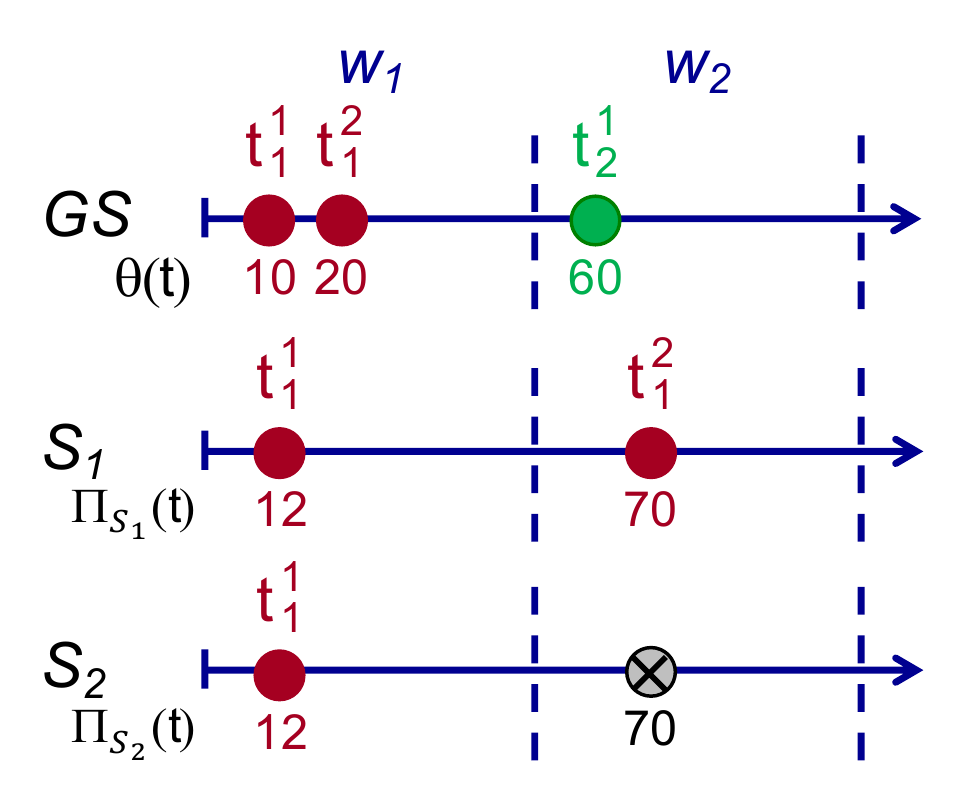}
    \captionof{figure}{Examples of runs retrieved by $S_1$ and $S_2$ as well as the associated ground truth ($GS$) with respect to H1. In this example, a time window $w_i$ lasts 50 seconds. }
 \label{fig:exampleH1} 
\end{minipage}\hspace{10px}
\begin{minipage}[b]{.55\textwidth}\centering
{\tiny
\begin{tabular}{lp{0.42\textwidth}p{0.48\textwidth}}\toprule
Metrics         & \multicolumn{2}{c}{Systems}\\\cmidrule{2-3}
         & \multicolumn{1}{c}{$S_1$} & \multicolumn{1}{c}{$S_2$}\\\midrule
EG-0         &  $(\frac{1}{2}*1+0)/2=0.25$ & $(\frac{1}{1}*1+0)/2=0.5$ \\[0.10cm]
EG-1         & $(\frac{1}{2}*1+0)/2=0.25$ & $(\frac{1}{1}*1+0)/2=0.5$ \\[0.10cm]
EG-p        & $(\frac{1}{2}*1+0)/2=0.25$ & $(\frac{1}{1}*1+0)/2=0.5$ \\[0.10cm]\midrule[0.03em]
nCG-0        & $(\frac{1}{1}*1+\frac{1}{1}*0)/2=0.5$ & $(\frac{1}{1}*1+0)/2=0.5$ \\[0.10cm]
nCG-1        & $(\frac{1}{1}*1+0)/2=0.5$ & $(\frac{1}{1}*1+0)/2=0.5$ \\[0.10cm]
nCG-p       & $(\frac{1}{1}*1+0/2=0.5$ & $(\frac{1}{1}*1+0)/2=0.5$ \\[0.10cm]
\midrule[0.03em]
GMP.50       & $((0.5*1-0.5*1)+(0))/2=0$ &  $((0.5*1)+(-0.5*1))/2=0$ \\[0.10cm]
\midrule[0.03em]
Latency     & $2$ & $2$\\\bottomrule
\end{tabular}}
\captionof{table}{Behaviours of the studied metrics with respect to H1.}
\label{tab:exampleH1}
\end{minipage}
\vspace{-0.5cm}
\end{figure}

\noindent{\bf H1.} Considering the example of Fig. \ref{fig:exampleH1} and Table \ref{tab:exampleH1}, the system $S_2$ has higher scores than $S_1$ on the \textit{EG} metrics, whereas both return results supposed as equivalent (the first tweet of the cluster $C_1$  during $w_1$ and respectively a redundant and non relevant tweet during $w_2$). $S_1$ is more penalized for returning a redundant tweet than a non relevant one. This thus violates {\bf H1}.

\begin{figure}[H]
\begin{minipage}[b]{.35\textwidth}\centering
  \includegraphics[width=0.8\textwidth]{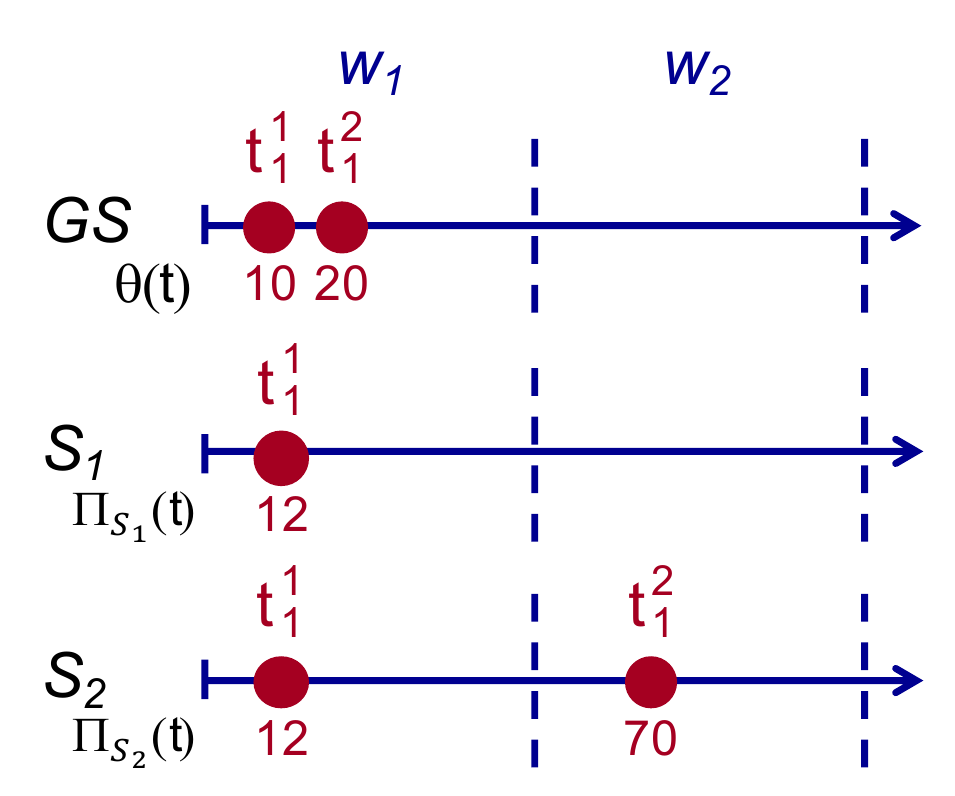}
    \captionof{figure}{Examples of runs retrieved by $S_1$ and $S_2$ as well as the associated ground truth ($GS$) with respect to H2. In this example, a time window $w_i$ lasts 50 seconds.}
 \label{fig:exampleH2} 
\end{minipage}\hspace{10px}
\begin{minipage}[b]{.55\textwidth}\centering
{\tiny
\begin{tabular}{lp{0.42\textwidth}p{0.48\textwidth}}\toprule
Metrics         & \multicolumn{2}{c}{Systems}\\\cmidrule{2-3}
         & \multicolumn{1}{c}{$S_1$} & \multicolumn{1}{c}{$S_2$}\\\midrule
EG-0         &  $(\frac{1}{1}*1+0)/2=0.5$ & $(\frac{1}{2}*1+0)/2=0.25$ \\[0.10cm]
EG-1         & $(\frac{1}{1}*1+1)/2=1$ & $(\frac{1}{2}*1+1)/2=0.75$ \\[0.10cm]
EG-p        & $(\frac{1}{1}*1+1)/2=1$ & $(\frac{1}{2}*1+1)/2=0.75$ \\[0.10cm]\midrule[0.03em]
nCG-0        & $(\frac{1}{1}*1+0)/2=0.5$ & $(\frac{1}{1}*1+0)/2=0.5$ \\[0.10cm]
nCG-1        & $(\frac{1}{1}*1+1)/2=1$ & $(\frac{1}{1}*1+1)/2=1$ \\[0.10cm]
nCG-p       & $(\frac{1}{1}*1+1)/2=1$ & $(\frac{1}{1}*1+1)/2=1$ \\[0.10cm]
\midrule[0.03em]
GMP.50       & $((0.5*1)+(0))/2=0.25$ &  $((0.5*1-0.5*1)+(0))/2=0$ \\[0.10cm]
\midrule[0.03em]
Latency     & $2$ & $2$\\\bottomrule
\end{tabular}}
\captionof{table}{Behaviours of the studied metrics with respect to H2.}
\label{tab:exampleH2}
\end{minipage}
\end{figure}

\noindent{\bf H2.}
Considering the example of Fig.  \ref{fig:exampleH2} and Table \ref{tab:exampleH2}, $w_2$ is a silent day for both systems. $S_2$ breaks the silence with $t_1^2$ and however obtains a perfect score on this day, as $S_1$ which did not push any tweet. This thus violates {\bf H2}.

These two counterexamples are a side effect of {\bf S2}.

%% file: sections/discussion.tex
\section{Discussion of the RTS Settings}
\label{sec:hyperparameters}
\begin{figure}[H]
\centering
\includegraphics[width=0.5\textwidth]{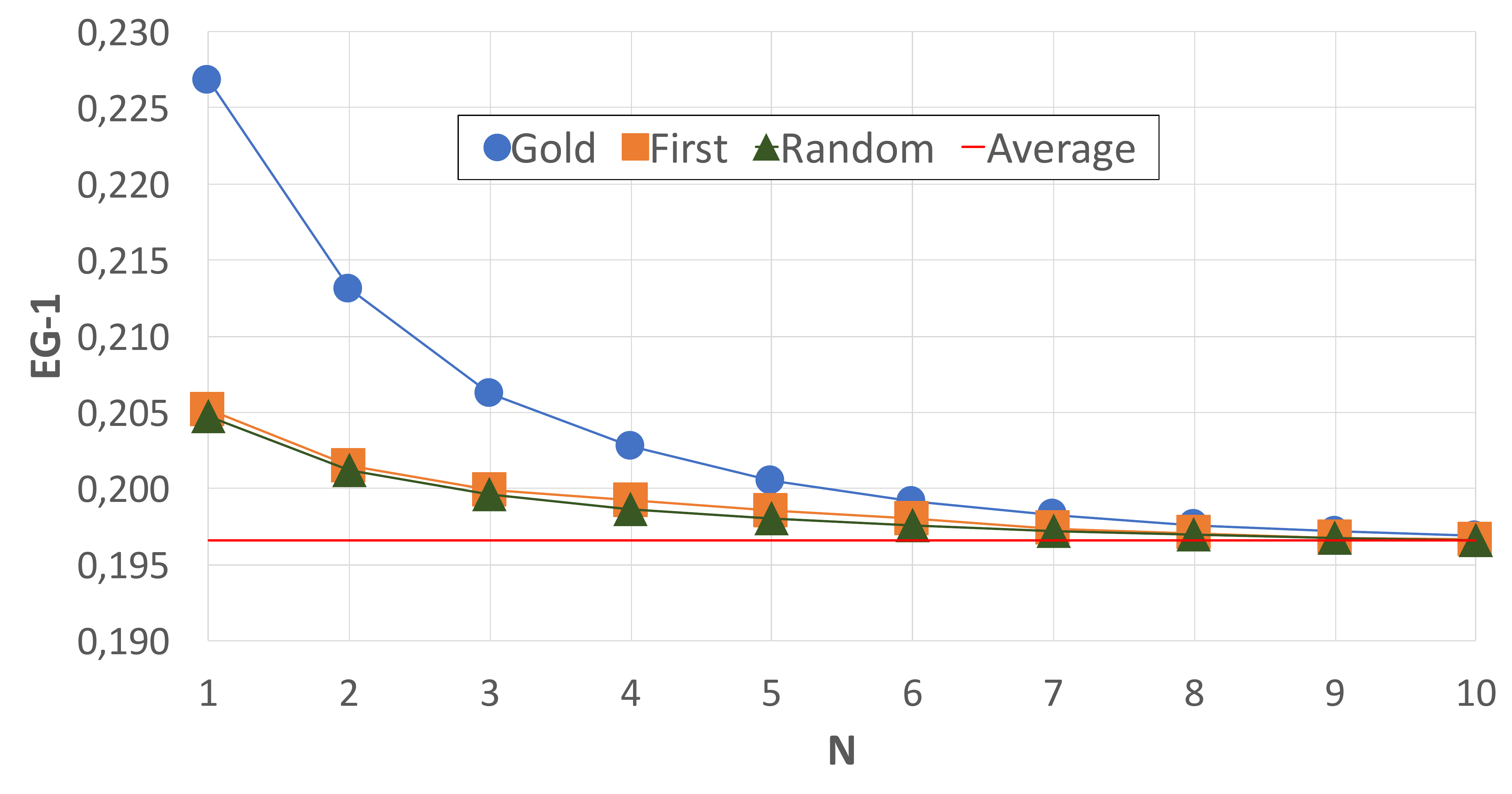}
\vspace{-0.25cm}
\caption{Impact of S1 on the \textit{EG}-1 metric}
\label{fig:k}
\end{figure}

\noindent{{\bf S1.}} Allowing up to 10 tweets to be pushed per profile per day is an arbitrary limit of the task. In this section, we wonder how much the 2016 official metric would have been impacted by a modification of this value.  With this aim in mind, we adopt the following methodology. Given $N \in \{1..10\}$, we apply three distinct strategies to restrict the 2016 official runs to push only $N$ tweets per profile per day and then calculate the average value of the \textit{EG}-1 metric\footnote{At the time of the paper submission, the 2017 official runs are not available.}. The three strategies are as follows:
\begin{itemize}
\item In the {\sc First} strategy, the first $N$  tweets according to their pushing date are considered. This strategy intuitively simulates a change in the setting but no self-adaptation of the systems to this tighter constraint.
\item In the {\sc Gold} strategy, $N$ tweets are chosen to maximize the number of clusters and thus the official metric. Given a window $w_j$, a profile $p$, and a system $S_i$, if $N$ is greater than the number of clusters retrieved by $S_i$ during $w_j$ for $p$, non-relevant tweets, i.e., either redundant or irrelevant tweets, are pushed to fulfill our requirement. Contrary to the {\sc First} strategy, this strategy simulates a self-adaptation of the systems under this tighter constraint.  
\item In the {\sc Random} strategy, $N$ tweets are randomly chosen. To overcome any bias in the sampling, 100 random draws were performed and the \textit{EG}-1 metric values for all these 100 runs were then averaged. This strategy represents a fair compromise between the naive {\sc First} and the optimal {\sc Gold} strategies.
\end{itemize}
It should be noted that if less than $N$ tweets have been pushed by a system $S_i$ for a profile $p$ during a window $w_j$, the set returned by any strategy is the same as the original set of pushed tweets. Finally, to fairly evaluate the impact of varying $N$, we compare the results obtained by these strategies to the average value of \textit{EG}-1 in the official runs. The impact of the window size on the aforementioned strategies is shown in Fig.~\ref{fig:k}. Several strong conclusions can be drawn from these results. First, whatever the strategy, it is always beneficial to return very few tweets reinforcing the idea that \textit{EG}-1 is essentially a precision-oriented metric. This conclusion is obviously even more true for the  {\sc Gold} strategy. Second, the performances of the {\sc First} and {\sc Random} strategies are very close suggesting that relevant tweets retrieved by the systems are uniformly distributed over the time window. Last but not least, pushing only one tweet adopting a very basic strategy, i.e., either the {\sc Random} or the {\sc First} one, without any guarantees that this tweet is relevant,  provides similar or better results than pushing 3 or more tweets using a sophisticated strategy such as the {\sc Gold} one, i.e., in which the number of retrieved clusters is maximized. This very interesting result reinforces our claim about the regrettable non-consideration of the coverage in the official metric. This point will be further discussed in Section~\ref{sec:reco}.

\noindent{{\bf S2.}} While the side effect of S2 have been assessed in Section~\ref{sec:metrics}, we now focus on its practical consequences during the TREC RTS 2016 track. For this purpose, we simulated an alternative evaluation framework in which the tweets \emph{are not} sent back to their publication window. 

 We observe from these statistics that:
\begin{itemize}
\item very few tweets have been pushed in another window than their creation window. Specifically, this concerns only $0.12\,\%$ ($53/45751$) of the pushed tweets all the systems taken together.
\item these $53$ tweets have been pushed by $8$ different systems over the $41$ runs. Notably, one of these systems have pushed $42$ tweets among the $53$ tweets while the other $7$ systems have pushed only $1$ or $2$ tweets outside their creation window.
\item due to the rarity of these push window gaps, there is no differences in the rankings whatever the date taken into account for the evaluation.
\end{itemize}    

Note that S2 could theoretically impact the performances of the systems, but this situation is not observed in the 2016 runs. Moreover, S2 may send back tweets to a window without any restrictions and making it greater than $N$, calling into question S1.

%% file: sections/rerun.tex
\section{Reusability of the Test Collection}\label{sec:rerun}

In order to enable comparison of new solutions against the TREC RTS 2016 results, the organizers publicly provide an evaluation script as well as 3 ground truth files: (i) the \textit{qrels} file that contains the relevance level of each tweet from the pool, (ii) the \textit{cluster} file that gives cluster for each profile and (iii) the  \textit{epoch} file that contains the publication date of tweets from the pool.

We conducted a standard ``leaved-one out'' analysis to evaluate the reusability of the collection. To do so, 
we simulated a rerun setup for all the 41 runs submitted during 2016 and evaluated them using the official metric, \textit{EG}-1. The ground truth files, i.e., cluster, qrels, and epoch files, were created for each run$_i$ as if it has not taken part in the track by removing its unique tweets. For each of these new 41 evaluation files, an alternative ranking$_i$ was obtained using the \textit{EG}-1 metric. The official ranking of each run$_i$ was then compared to this new ranking$_i$ in order to determine how effective would have been this run$_i$ in a rerun setup. The position of each run$_i$ in the ranking$_i$ showed either improvement or no variability with respect to its position in the original ranking, resulting in an average gain of 2.1 positions. This very surprising result has motivated a deep analysis of the evaluation tool. We observed a very odd behavior on how the unassessed tweets, i.e., the tweets that are not referenced in the ground truth files, are considered. Indeed, such tweets are simply ignored instead of being considered as irrelevant as traditionally done in classical evaluation setups. This point is even more problematic since the way the runs deal with the silent days is crucial for the calculation of gain-oriented metrics. By decreasing the number of tweets per profile/day and increasing chances to respect the silent days, the performances of new runs in the rerun setup are artificially increased.
This situation is only attenuated, but still not solved, thanks to the introduction of the \textit{EG}-p and \textit{nCG}-p metrics in 2017. However, ignoring the tweets that must be considered as not relevant will still increase the score obtained by those metrics during the silent days. 
This bias in the (re)evaluation can be solved by including all the tweets of the Twitter stream during the evaluation period (11.5\,M tweets) in the epoch file. In this case, our results showed a different ranking behaviour. None of the runs improved its position in its respective ranking, dealing with an average lost of 0.6 positions when compared to their original position.

Regarding the settings, we would like to draw attention to the fact that S1 is under the responsibility of each user of this collection. This setting was automatically handled by the organizers through the broker during the task. Not respecting this limit during the rerunning leads to underestimated performance since the gain is calculated only over the first 10 tweets but it is normalized by the total number of tweets, which could be greater than 10. Contrary to S1, S2 is always applied without user intervention. 

Finally, users of the collection must consider analysis and remarks presented in Sections \ref{sec:hypotheses} and \ref{sec:hyperparameters} because they are also valid under the rerun setup. We confirm the reusability of the collection only under the aforementioned conditions, in particular, use of a complete \textit{epoch} file and strict application of S1.

%% file: sections/reco.tex
\section{Recommendation and Conclusion}
\label{sec:reco}

To conclude, we would like to summarize our main findings in this paper:
\begin{itemize}
    \item we clarified some definitions and assumptions of the track guidelines. We highlight here two of them, which are not clearly stated in the guidelines and overwiews of the track although  crucial for a good understanding of the evaluation framework.  Only a deep analysis of the evaluation tool lead us to these conclusions, causing us to believe that some participants may not be conscious of these findings:
    \begin{itemize}
        \item the evaluation window used in EG and nCG metrics is \textit{not} the window corresponding to the tweet push-timestamp. Each returned tweet is sent back to its emission window, which significantly impacts the way metrics are evaluated.
        \item  \textit{silent days are system-dependent}. This is thus non-sense to elaborate approaches that try to detect silent days independently of already returned tweets.
    \end{itemize}
    \item we shed the light on the fact that \textit{coverage is not really evaluated by the official metrics}. The systems would better return few tweets that are very likely relevant to optimize the metrics. Trying to maximize the coverage and thus returning many tweets will probably lead to a result degradation. \textit{As a consequence, when developing a system for the track, all the improvements against the metrics  should be compared to a very simple run returning at most one tweet per time window}.
This behavior of the results has already been noticed by the track organizers \cite{QianSigir2016}, but this was credited to misconfigurations of the systems that returned very few tweets. On the contrary, we do think that, given the metrics and the way the silent days are considered,  systems should return few tweets to be top-ranked. This unusual behavior of the metrics is not observed on the other traditionally-used precision-oriented metrics such as P@K and MAP. 
Our official results on the 2017 track confirm these findings. We submitted a baseline run returning the first tweet of the day containing all the query terms (i.e., at most one tweet per profile and per day was returned). This very simple baseline allowed us to be ranked $2^{nd}$ on the mobile evaluation and $4^{th}$ on the batch one of Scenario A~\cite{overview2017}. 
\item Concerning the reusability of the collection, we found a problem on the \textit{epoch file} used in evaluation. In case of rerun, researchers should add all their tweets to the official epoch file, which is not mentioned in the evaluation tool documentation. Otherwise, the results are largely over-evaluated since the evaluation does not consider the non-relevant tweets that are absent from the epoch file. As this problem has never been mentioned before by track organizers or participants,  it is very likely that some already-published research papers using the TREC RTS collection as evaluation framework report over-evaluated results.  \\

\end{itemize}

In future and concerning the metrics, since the track will be pursued in 2018, we suggest to focus on the relative importance of clusters. For instance, let us consider the 2017 profile RTS60 entitled ``Beyonce's babies".
The very famous photo posted on Instagram in which Beyonce officially announced the names of the twins with their first image is a crucial information for this profile. Other information such as the name of the nurse is also relevant but less crucial.

With equal numbers of retrieved clusters, the systems that find the cluster about this first announcement should thus be more rewarded than the systems that do not find it.

Separating latency and effectiveness should also be (re)considered. In 2015 the TREC microblog track included a very first  version of the task (named \textit{Scenario A - Push notification}) where a latency penalty was applied to the \textit{EG} metric~\cite{lin2015}. The metric has been given up since 2016 to understand the potential tradeoffs between quality and latency. However, we think that separating latency and gain metrics may lead to some side effects that could be avoided  with a single-point metric.